\documentclass[%
 reprint,
 amsmath,amssymb,
 aps,
floatfix,
]{revtex4-1}

\usepackage{graphicx}
\usepackage{dcolumn}
\usepackage{bm}

\usepackage{color}
\usepackage{soul}
\usepackage{ulem}

\begin{document}

\preprint{APS/123-QED}

\title{Continuous supersymmetric transformations in optical waveguides}

\author{G. Queralt\'{o}}
\email{Gerard.Queralto@uab.cat}
\author{V. Ahufinger}
\author{J. Mompart}
\affiliation{Departament de F\'{\i}sica, Universitat
Aut\`{o}noma de Barcelona, E--08193 Bellaterra, Spain}%

\date{\today}

\begin{abstract}
We introduce continuous supersymmetric transformations to manipulate the modal content in systems of optical waveguides, providing a systematic method to design efficient and robust integrated devices such as tapered waveguides, single-waveguide mode filters, beam splitters and interferometers. These transformations connect superpartner profiles by smoothly modifying the transverse index profile along the propagation direction and, if the modification is performed adiabatically, the transverse electric modes evolve adapting their shape and propagation constant without being coupled to other guided or radiated modes. Numerical simulations show that very high fidelities are obtained for a broad range of devices lengths and light's wavelengths.

\end{abstract}

\maketitle


\section{\label{sec:1}Introduction}

Photonic integrated devices offering high fidelity, high speed transmissions and scalability \cite{Lifante2003} have become very relevant in areas like optical communications \cite{Agrell2016}, lab-on-a-chip experiments \cite{Fan2008} or quantum technologies \cite{Meany2015}. Therefore, the development of new techniques to design integrated devices, such as tapered waveguides \cite{Heilmann2018}, photonic lanterns \cite{Birks2015}, mode filters and multiplexers \cite{Dai2017},  y-junctions \cite{Love2012} or interferometers \cite{Schipper1997}, with enhanced performances is of the main interest. One of the most recent proposals to this aim has been the application of Supersymmetry (SUSY), discovered in the 70's and applied to many areas of Physics \cite{Dine2007}, to optical systems by exploiting the analogies between the Schr\"odinger \cite{Cooper1995} and Helmholtz equations \cite{Chumakov1994}. In guided wave optics, SUSY establishes global phase-matching conditions among the modes of two different structures, called superpartners, except for the fundamental mode, bringing new opportunities for mode filtering and multiplexing \cite{Miri2013-2,Heinrich2014-1,Principe2015,Gerard2017,Macho2018}. In addition, SUSY techniques have also been applied to design refractive index profiles with nontrivial properties \cite{Miri2013-1,Miri2014}, systems with identical scattering characteristics \cite{Heinrich2014-2,Longhi2015} or digital multimode devices \cite{Yu2017}. 

So far, SUSY-based optical devices have mainly been applied to evanescently coupled modes of discrete superpartner structures. Here, instead, we consider a structure where the transverse index profile is adiabatically modified along the propagation direction such that, at the input and output ports, one has superpartner index profiles. In the most general case of a photonic lattice with $N$ waveguides, its superpartner has $N-1$ dissimilar channels \cite{Miri2013-2,Heinrich2014-1}, and, by connecting both profiles, one is able to design structures with different number of channels at the input and output ports. In particular, continuous SUSY transformations offer a systematic way to create tapered waveguides and mode filters using a single-waveguide structure or beam splitters and interferometers using a two-waveguide structure.

Our manuscript is organized as follows. In Sec.~\ref{sec:2} we develop the theoretical model describing the continuous SUSY transformation of a given refractive index profile along the propagation direction and discuss the adiabaticity conditions. In Sec.~\ref{sec:3} we describe different optical devices designed by applying the described technique. In Sec.~\ref{sec:4}, we present the results obtained through numerical simulations. Finally, in Sec.~\ref{sec:5}, we conclude and discuss future perspectives.

\section{\label{sec:2}Theoretical model}

\begin{figure}[t!]
\centering
\includegraphics[width=3.3 in]{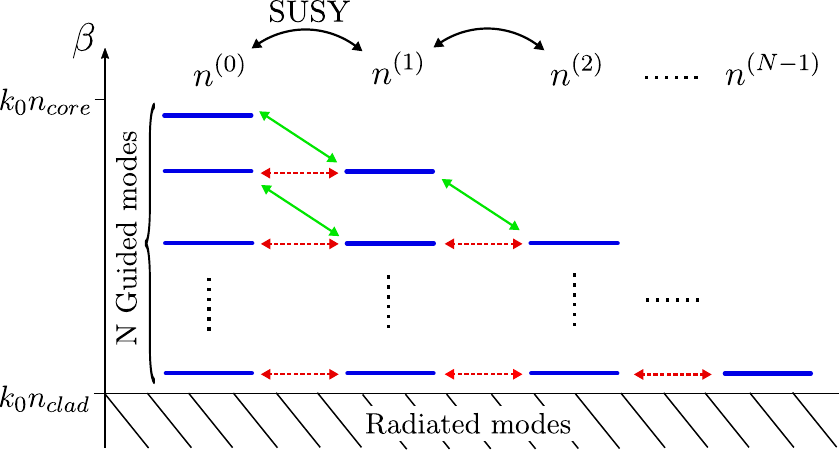}
\caption{Schematic representation of a hierarchical sequence of superpartner structures. The diagonal arrows indicate the evolution of the modes when the index profile is adiabatically modified along the propagation direction connecting the superpartner profiles. The horizontal arrows show the resonant couplings of the modes between discrete superpartners.}
\label{fig:1}
\end{figure}

In the paraxial approximation, the propagation of the transverse electric $(TE_m)$ component of the electric field along $z$-direction through a medium with arbitrary index of refraction $n(x,z)$ is described by the Helmholtz equation $\{\nabla^2+[k_0 n(x,z)]^2 \}E_y(x,z)=0$, where $k_0=2\pi/\lambda_0$ is the vacuum wavenumber. The electric field can be expressed as a superposition of modes as \cite{Love1983}: 
\begin{equation}
E_y(x,z)=\sum_{m} a_m(z) e_m(x,z) \exp\left[i\int_0^z{\beta_m(z) dz}\right],
\label{eq:Efield}
\end{equation}
where $a_m(z)$ is the amplitude, $e_m(x,z)$ the transverse spatial distribution, and $\beta_m(z)$ the propagation constant of  mode $m$. At any fixed position along the propagation direction, the problem is described by the eigenvalue equation $\mathcal{H}e_{m}(x)=\beta_{m}^2 e_{m}(x)$, where $\mathcal{H}=d^2/dx^2+[k_0 n(x)]^2$. The first superpartner profile $n^{(1)}(x)$ of a given $n^{(0)}(x)$ can be obtained, as long as the fundamental $(m=0)$ mode of the system is node free, by factorizing $\mathcal{H}$ following discrete SUSY techniques \cite{Miri2013-2}:
\begin{equation}
n^{(1)}(x)=\frac{1}{k_0}\sqrt{\left(\beta_0^{(0)}\right)^2-\left(W^{(0)}\right)^2 - \frac{dW^{(0)}}{dx}},
\label{eq:index-of-refraction}
\end{equation}
where $W^{(0)}(x)=-\partial_x\ln e_0^{(0)}(x)$ is the so-called superpotential. These transformations can be applied iteratively allowing the systematic design of superpartner profiles, as it is shown in Fig.~\ref{fig:1}. In what follows, the number of discrete SUSY transformations applied will be identified by the superscript $q$.

To achieve a continuous SUSY transformation connecting two superpartner profiles $n^{(q)}(x)$ and $n^{(q+1)}(x)$, we propose to smoothly modify the transverse refractive index profile along the propagation direction by introducing a continuous transformation function $g_{q}(z)$, valued between $0$ and $1$, from $z=L_q$ to $z=L_{q+1}$. The index profile of the structure is then characterized by:
\begin{equation}
n^{(q)\rightarrow (q+1)}(x,z)=\sqrt{\left(n^{(q)}(x)\right)^2-g_{q}(z)\frac{2}{k_0^2} \frac{dW^{(q)}}{dx}}.
\label{eq:index-of-refraction-2}
\end{equation}
To perform the modification adiabatically leading to the evolution of the propagating $m$ mode that follows the diagonal arrows in Fig.~\ref{fig:1}, one should avoid the coupling to other guided or radiated modes. To avoid coupling to any guided $l$ mode, the following adiabaticity condition should be satisfied:
\begin{equation}
\left|\langle e_l | \frac{d e_m}{dz} \rangle \right| \ll |\beta_m(z)-\beta_l(z)|,
\label{eq:adiabaticity1}
\end{equation}
where $\langle e_l | \frac{d e_m}{dz} \rangle$ denotes the corresponding overlap integral. For symmetric variations of the refractive index profile along $z$, $\langle e_{l} | \frac{d e_m}{dz} \rangle = 0$ between modes with opposite parity. To avoid coupling with the radiated modes, the adiabaticity condition to be fulfilled is:
\begin{equation}
\left|\langle e_{rad} | \frac{d e_m}{dz} \rangle \right| \ll |\beta_m(z) - k_0 n_{clad}|,
\label{eq:adiabaticity2}
\end{equation}
where $k_0n_{clad}$ fixes the minimum propagation constant above which the modes start to be radiated. In the following sections, we will discuss the specific shapes of $g_q(z)$ required to fulfill the adiabaticity conditions for  different photonic devices. 

\begin{figure*}[t!]
\centering
\includegraphics[width=6.7 in]{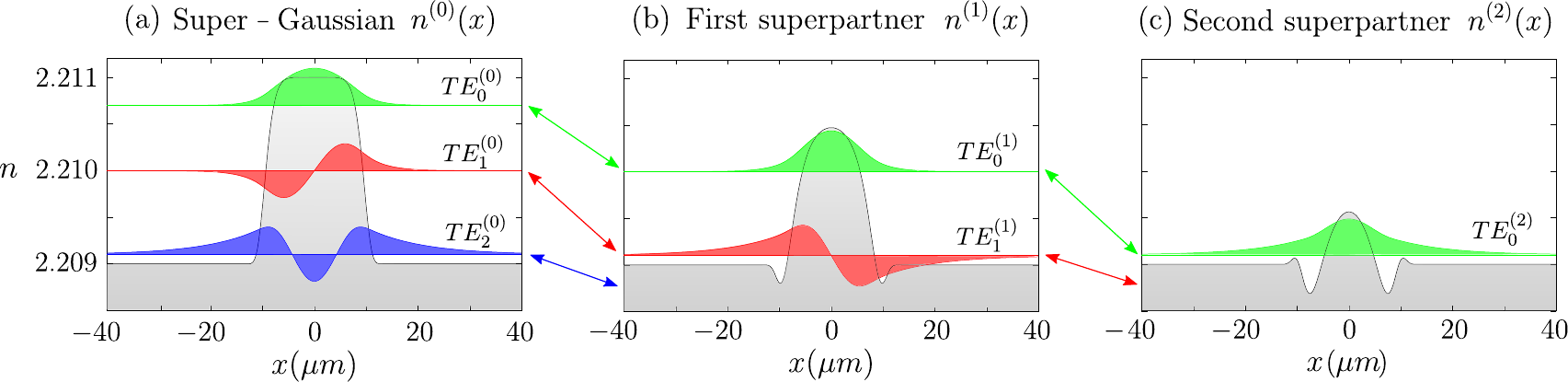}
\caption{Refractive index distribution and transverse mode amplitudes of (a) a Super-Gaussian profile $n^{(0)}(x)$ with $d=20 \rm{\mu m}$ and $2p=8$, (b) its first superpartner profile $n^{(1)}(x)$, and (c) its second superpartner profile $n^{(2)}(x)$. The positions of the modes along the vertical axis correspond to $\beta_m/k_0$ and the arrows indicate the evolution of the modes when the index profile is adiabatically modified along the propagation direction.}
\label{fig:2}
\end{figure*}

\section{\label{sec:3}Physical system}

Two configurations will be investigated: (i) a single-waveguide structure offering a systematic way to design tapered waveguides, used to propagate modes between waveguides with different widths, to avoid single-photon loss due to mode profile mismatch \cite{Heilmann2018} or to filter higher order modes by radiating them, and (ii) a two-waveguide structure which allows to design a beam splitter and a Mach-Zehnder Interferometer (MZI). The latter can be used to detect a very small variation of the refractive index giving rise to a differential phase shift that modulates the output intensity \cite{Fan2008}. 

To be specific, we will consider the parameters of $LiNbO_3$ waveguides, known for its high electro-optic coefficient \cite{Weis1985}, with refractive indices $n_{core}=2.111$ and $n_{clad}=2.209$ at telecom wavelength $\lambda=1.55 \, \rm{\mu m}$. Although we have focused on a low-contrast index structure corresponding to the state-of-the-art parameters in the experimental realization of SUSY waveguides \cite{Heinrich2014-1}, the results are not limited to these refractive index contrast and profiles \cite{Miri2014-3}.

\subsection{Single-waveguide structure}

\begin{figure}[t!]
\centering
\includegraphics[width=2.2 in]{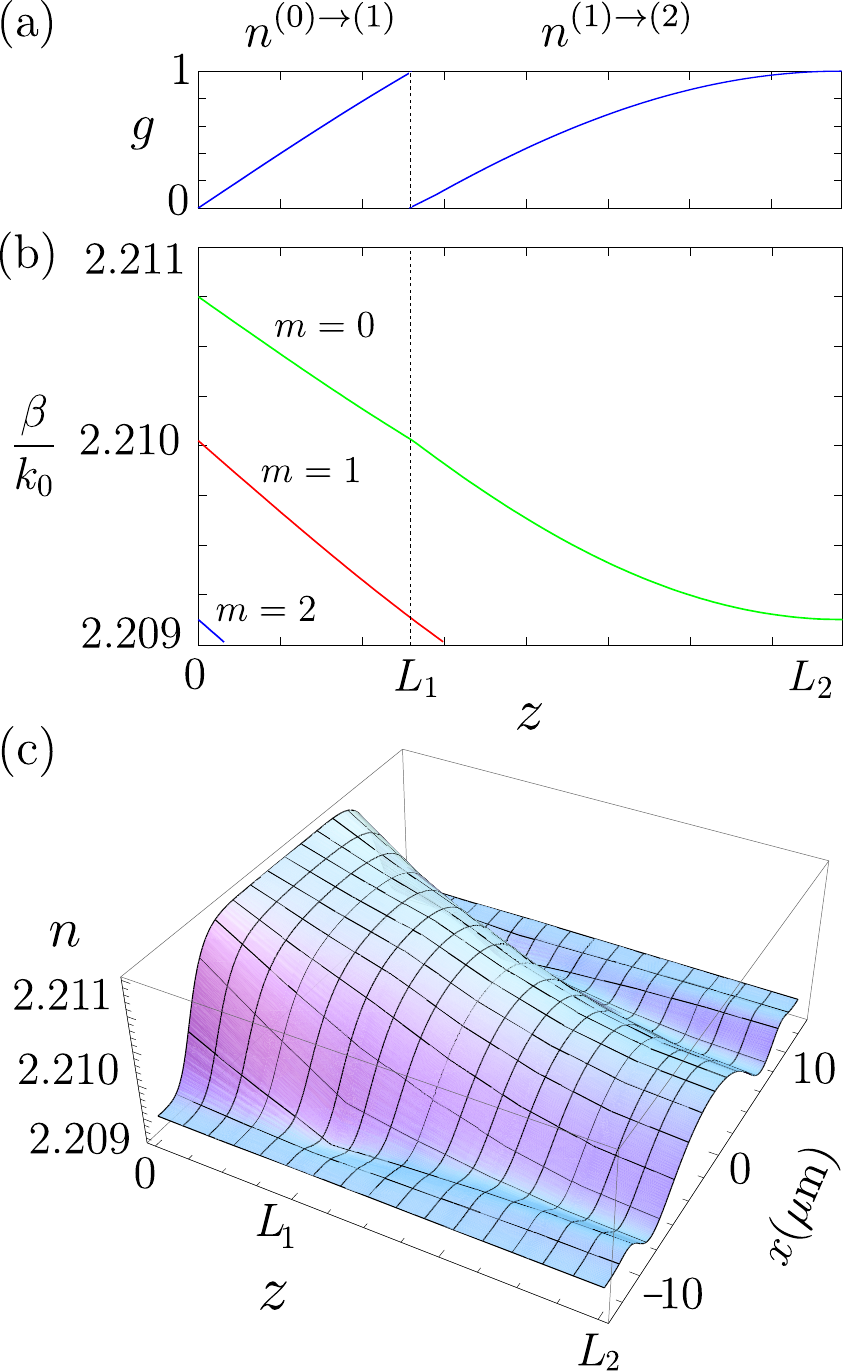}
\caption{Evolution along the propagation direction for the single-waveguide case of (a) the continuous transformation function $g_q(z)$, (b) the propagation constants $\beta_m(z)$, and (c) the refractive index profile corresponding to the $n^{(0) \rightarrow (1)}(x,z)$ transformation between $0 \leq z \leq L_1$ and $n^{(1) \rightarrow (2)}(x,z)$ transformation between $L_1 \leq z \leq L_2$.}  
\label{fig:3}
\end{figure}

We consider a waveguide whose index of refraction is defined by a super-Gaussian profile: 
\begin{equation}
\left[n^{(0)}(x)\right]^2=n_{clad}^2+(n_{core}^2-n_{clad}^2)e^{\left[-\left(\frac{2x}{d}\right)^{2p}\right]},
\label{eq:sg}
\end{equation}
where $2p$ is an index that smoothes the profile \cite{Miri2014-3} and the width $d$ of the waveguide is selected to allow the propagation of the $TE_0^{(0)}$, the $TE_1^{(0)}$ and the $TE_2^{(0)}$ modes with propagation constants $\beta_{0}^{(0)}$, $\beta_1^{(0)}$, and $\beta_2^{(0)}$, respectively (see Fig.~\ref{fig:2}(a)). By applying discrete SUSY techniques, one obtains $n^{(1)}(x)$ supporting the $TE_0^{(1)}$ and the $TE_{1}^{(1)}$ modes with $\beta_0^{(1)}=\beta_1^{(0)}$ and $\beta_1^{(1)}=\beta_2^{(0)}$ (see Fig.~\ref{fig:2}(b)), and $n^{(2)}(x)$ supporting only the $TE_0^{(2)}$ mode with $\beta_0^{(2)}=\beta_1^{(1)}=\beta_2^{(0)}$ (see Fig. \ref{fig:2}(c)). 

Considering the $n^{(0)\rightarrow (1) \rightarrow (2)}(x,z)$ transformation with $n^{(0)}(x)$ as the input and $n^{(2)}(x)$ as the output ports, if the transverse refractive index profile is modified along the propagation direction fulfilling the adiabaticity conditions, the propagating modes will evolve adapting their shape and propagation constant as indicated by the arrows of Fig.~\ref{fig:2}. In particular, if the $TE_0^{(0)}$ mode is injected, it will be converted into the $TE_0^{(2)}$ mode at the output port and the device will work as an efficient tapered waveguide. For parity reasons, the $TE_0^{(0)}$ mode can only be coupled to the $TE_2^{(0)}$ mode, and, since the latest will become a radiated mode after a short propagation distance, it is enough to fulfill Eq.~(\ref{eq:adiabaticity2}) for the fundamental mode to avoid the coupling with the radiated modes. Therefore, we propose to use $g_q(z)=4\cos^2(A_q(z))-B_q$, where $A_q(z)$ and $B_q$ are used to bound the values of the function between $0$ and $1$. For the $n^{(0)\rightarrow (1)}(x,z)$ transformation occurring between $L_0=0\leq z \leq L_1$, we use $A_0(z)=\frac{\pi}{12}\frac{z}{L_1}+\frac{3\pi}{4}$ and $B_0=2$, being $g_0(z)$ approximately linear, while for the $n^{(1)\rightarrow (2)}(x,z)$ transformation occurring between $L_1 < z \leq L_2$, we use $A_1(z)=\frac{\pi}{6}\frac{(z-L_1)}{(L_2-L_1)}+\frac{5\pi}{6}$ and $B_1=3$ with $g_1(z)$ becoming smoother as it approaches $z=L_2$ (see Fig.~\ref{fig:3}(a)). The evolution of the propagation constants and the refractive index profile along the $z$-direction are shown in Fig.~\ref{fig:3}(b) and Fig.~\ref{fig:3}(c), respectively. 

Furthermore, the $n^{(0)\rightarrow (1) \rightarrow (2)}(x,z)$ structure could be used as a single-waveguide mode filter since, if a superposition of the $TE_0^{(0)}$, the $TE_1^{(0)}$ and the $TE_2^{(0)}$ modes is injected through the input port, the $TE_1^{(0)}$ and the $TE_2^{(0)}$ modes will be radiated during the $n^{(1) \rightarrow (2)}(x,z)$ and $n^{(0)\rightarrow (1)}(x,z)$ transformations, respectively. 
 
\subsection{Two-waveguide structure}

\begin{figure*}[t!]
\centering
\includegraphics[width=6.5 in]{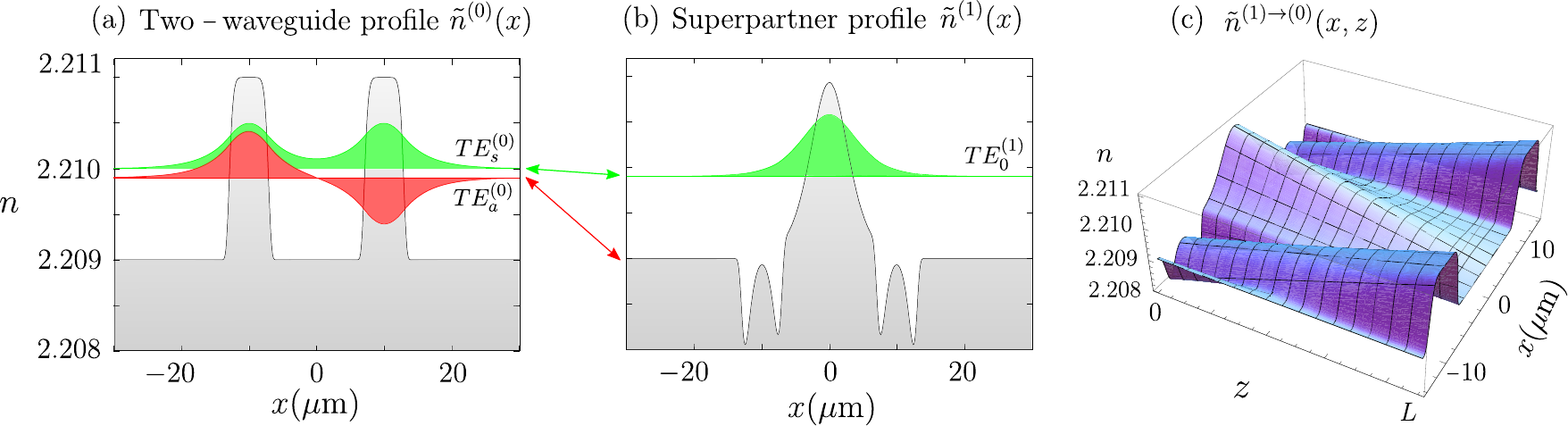}
\caption{Refractive index profile and transverse mode amplitudes of (a) a two-waveguide super-Gaussian profile $\tilde{n}^{(0)}(x)$ with $d=8 \, \mu \rm{m}$ and $D= 20 \, \mu \rm{m}$, and (b) its superpartner profile $\tilde{n}^{(1)}(x)$. The positions of the modes along the vertical axis correspond to $\beta_m/k_0$ and the diagonal arrows indicate the evolution of the modes when the index profile is modified adiabatically along the propagation direction. (c) Refractive index profile corresponding to the $\tilde{n}^{(1) \rightarrow (0)}(x,z)$ transformation between $0 \leq z \leq L$.}
\label{fig:4}
\end{figure*}

Now, we consider a two-waveguide structure $\tilde{n}^{(0)}(x)$ characterized by two identical evanescently-coupled waveguides separated a distance $D$, being each of them defined by the super-Gaussian profile of Eq.~(\ref{eq:sg}). The waveguides are single-mode in isolation and when they are coupled, the structure supports the symmetric $TE_s^{(0)}$ and the antisymmetric $TE_a^{(0)}$ supermodes with propagation constants $\tilde{\beta}_s^{(0)}$ and $\tilde{\beta}_a^{(0)}$, respectively (see Fig.~\ref{fig:4}(a)). By applying discrete SUSY techniques, one obtains the superpartner profile $\tilde{n}^{(1)}(x)$ supporting only the $TE_0^{(1)}$ mode with propagation constant $\tilde{\beta}_0^{(1)}=\tilde{\beta}_a^{(0)}$, as it is shown in Fig.~\ref{fig:4}(b). 

Considering the $\tilde{n}^{(1)\rightarrow (0)}(x,z)$ transformation, if the $TE_0^{(1)}$ mode is injected and the modification of the refractive index profile is performed adiabatically, it will evolve following the arrow of Fig.~\ref{fig:4} becoming the $TE_s^{(0)}$ supermode at the output port. For parity reasons, there is no coupling between the two guided modes and the restriction to have an adiabatic evolution is only given by Eq.~(\ref{eq:adiabaticity2}). Since the difference $|\beta_m(z) - k_0 n_{clad}|$ is approximately constant, a linear $\tilde{g}_0(z)$ fulfills the adiabaticity condition, obtaining the refractive index profile represented in Fig.~\ref{fig:4}(c). 

This configuration could be used as a symmetric beam splitter since the power injected through the input port will be divided $50\%$ at each waveguide of the output port. By exchanging the input and output ports, the $\tilde{n}^{(0)\rightarrow (1)}(x,z)$ structure allows to recombine the beams into a single channel. Finally, a MZI could be designed by combining $\tilde{n}^{(1)\rightarrow (0)}(x,z)$, a central region with the two-waveguide profile $\tilde{n}^{(0)}(x)$ and $\tilde{n}^{(0)\rightarrow (1)}(x,z)$. 

\section{\label{sec:4}Results and discussion}

In this section, we will demonstrate through numerical simulations using Finite Difference Methods the efficiency and robustness of the tapered waveguide, the single-waveguide mode filter and the symmetric beam splitter. We will also test the MZI by simulating the application of a voltage in one of the arms, which changes the refractive index and thus, modulates the phase. To obtain the fidelity of each device we will compute:
\begin{equation}
\mathcal{F}_m=\left|\langle e_{out} | e_{m} \rangle \right|^2, 
\label{eq:Figure-of-merit}
\end{equation}
where $|e_{out}\rangle$ is the transverse modal field distribution numerically obtained at the output port and $|e_{m}\rangle$ is the theoretically expected transverse modal field distribution at this port. 

\subsection{Tapered waveguide and mode filter}

\begin{figure}[h!]
\centering
\includegraphics[width=3.0 in]{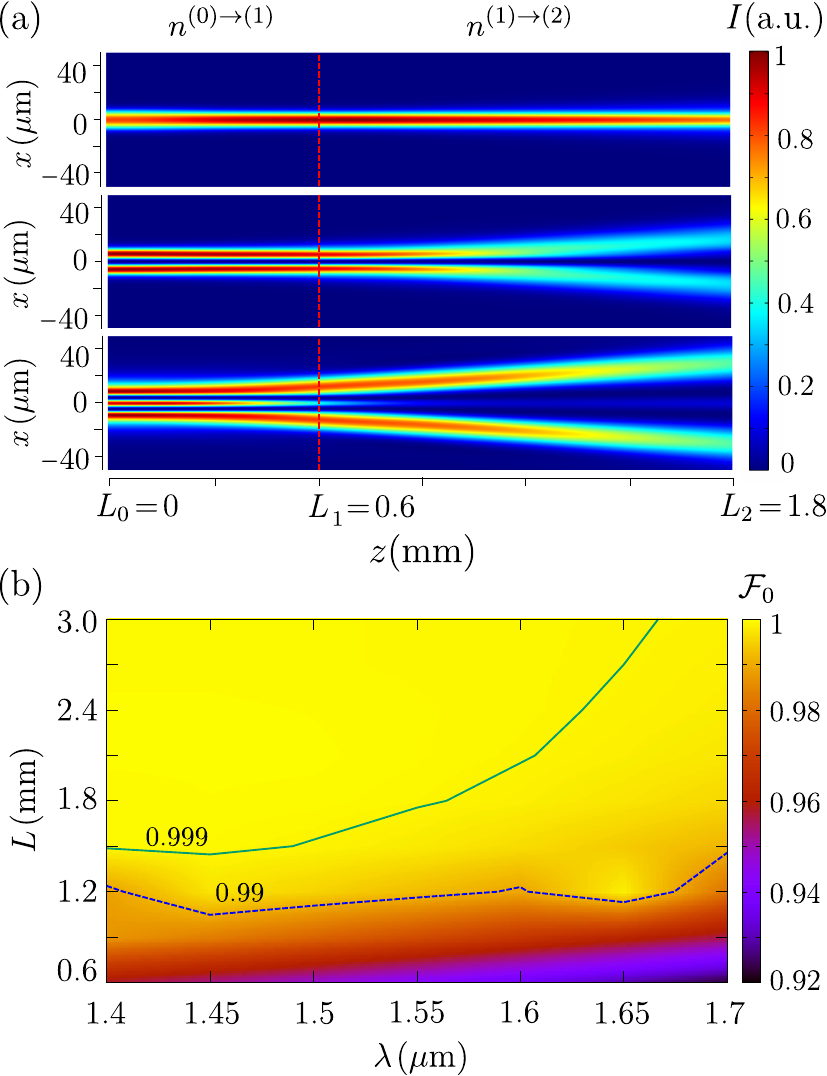}
\caption{(a) Numerical simulation of light intensity propagation ($\lambda=1.55 \, \mu \rm{m}$) along the $n^{(0)\rightarrow (1) \rightarrow (2)}(x,z)$ structure, see Fig.~\ref{fig:3}(c), when the $TE_0^{(0)}$ (upper panel), the $TE_1^{(0)}$ (middle panel), and the $TE_2^{(0)}$ (lower panel) mode is injected through the input port $n^{(0)}(x)$. The vertical dashed line delimits the two continuous SUSY transformations. (b) Fidelity of the tapered waveguide for the fundamental mode numerically calculated for different lengths and light's wavelengths.}
\label{fig:5}
\end{figure}

Here, we demonstrate the efficiency and robustness of the structure, characterized by the index profile transformation $n^{(0) \rightarrow (1) \rightarrow (2)}(x,z)$ illustrated in Fig.~\ref{fig:3}(c), working as a tapered waveguide and a mode filtering device. The numerical simulations confirm that, since the adiabaticity condition is fulfilled, the $TE_0^{(0)}$ mode is converted into the $TE_0^{(2)}$ mode at the output port, while the $TE_1^{(0)}$ and $TE_2^{(0)}$ modes become radiated modes during the continuous SUSY transformations. Figure \ref{fig:5}(a) shows the evolution of the $TE_0^{(0)}$ (upper panel), and the radiative loss of the $TE_1^{(0)}$ (middle panel) and the $TE_2^{(0)}$ (lower panel) modes along the propagation direction when they are injected through the input port $n^{(0)}(x)$. 

By using the proposed transformation functions $g_0(z)$ and $g_1(z)$ of Sec.~\ref{sec:3} and working at the telecom wavelength, we find that fidelities $\mathcal{F}_0 > 0.99$ and $\mathcal{F}_0 > 0.999$ are obtained for devices with total length $L > 1.2 \, \rm{mm}$ and $L > 1.8 \, \rm{mm}$, respectively (see Fig.~\ref{fig:5}(b)). Instead, if one uses a linear $g(z)$ function, to have $\mathcal{F}_0>0.999$ one would require $L > 10 \, \rm{mm}$. In addition, we prove the robustness of the device obtaining $\mathcal{F}_0 > 0.999$ in a broad region of wavelengths for devices with $L > 1.8 \, \rm{mm}$ (see Fig.~\ref{fig:5}(b)). For lower wavelengths the modes are more confined, less sensitive to the refractive index variations, and the fidelities are higher. Moreover, if one only uses the $n^{(0)\rightarrow (1)}(x,z)$ transformation, the device works as an efficient tapered waveguide for the $TE_0^{(0)}$ and the $TE_1^{(0)}$ modes, obtaining $\mathcal{F}_0 > 0.99$ and $\mathcal{F}_1 = 0.96$ , respectively, for $L_1=0.6 \, \rm{mm}$. Since the first excited mode is less confined and the $g_0(z)$ has not been optimized to fulfill Eq.~(\ref{eq:adiabaticity2}) for this mode, a longer device of $L_1 > 1.5 \, \rm{mm}$ would be required to obtain $\mathcal{F}_1>0.99$. 

In addition, we evaluate the fidelity $\mathcal{F}_{filter}=\mathcal{F}_0(1-\mathcal{F}_1)(1-\mathcal{F}_2)$ of the the $n^{(0) \rightarrow (1) \rightarrow (2)}(x,z)$ structure working as a mode filtering device. We obtain $\mathcal{F}_0=0.999$, $\mathcal{F}_1=0.245$, $\mathcal{F}_2=0.205$, yielding to $\mathcal{F}_{filter}=0.60$ for $L = 1.8 \, \rm{mm}$. To achieve fidelities above $0.9$, one could re-scale the $n^{(0) \rightarrow (1) \rightarrow (2)}(x,z)$ structure to $L>4 \, \rm{mm}$, prolongate the $n^{(2)}(x)$ profile of the output port along the propagation direction to spatially separate the modes, e.g. adding $1 \, \rm{mm}$ to the $L = 1.8 \, \rm{mm}$ device, or optimize the $g(z)$ function to faster radiate the modes without significantly decreasing $\mathcal{F}_0$. 

\begin{figure}[t!]
\centering
\includegraphics[width=3.0 in]{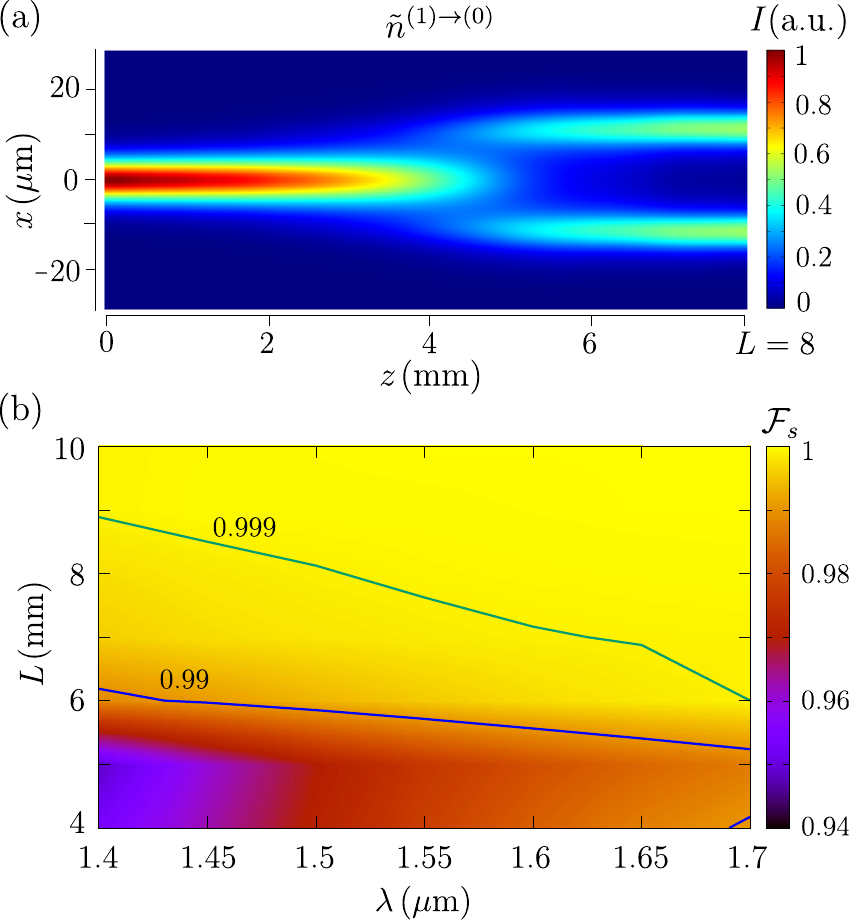}
\caption{(a) Numerical simulation of light intensity propagation ($\lambda = 1.55 \, \rm{\mu m}$) along the $\tilde{n}^{(1)\rightarrow (0)}(x,z)$ structure, see Fig.~\ref{fig:4}(c), when the $TE_0^{(1)}$ mode is injected through the input port $\tilde{n}^{(1)}(x)$. (b) Fidelity of the beam splitter calculated through numerical simulations for different lengths and light's wavelengths.}
\label{fig:6}
\end{figure}

\subsection{Beam splitter and MZI}

Here, we demonstrate the efficiency and robustness of the symmetric beam splitter, characterized by the index profile transformation $\tilde{n}^{(1) \rightarrow (0)}(x,z)$, represented in Fig.~\ref{fig:4}(c). Figure \ref{fig:6}(a) shows that, since the adiabaticity condition is fulfilled, the $TE_0^{(1)}$ mode injected through the input port $\tilde{n}^{(1)}(x)$ is converted into the $TE_s^{(0)}$ mode at the output port $\tilde{n}^{(0)}(x)$. Using a linear $g(z)$ and working at telecom wavelength, we find that the fidelity $\mathcal{F}_s=|\langle e_{out} | e_{s}^{(0)} \rangle |^2$ of the structure working as a beam splitter is $\mathcal{F}_s>0.99$ and $\mathcal{F}_s>0.999$ for $L > 6 \, \rm{mm}$ and $L > 8 \, \rm{mm}$, respectively. Fidelities $\mathcal{F}_s>0.99$ are obtained in a broad region of wavelengths for $L > 6 \, \rm{mm}$ confirming the robustness of the device, as can be seen in Fig.~\ref{fig:6}(b). This device resembles a symmetric y-junction, with similar fidelities for similar total length. Using waveguides with different widths, a device resembling an asymmetric y-junction could be designed and used for mode-division multiplexing applications \cite{Driscoll2013}.

\begin{figure}[t!]
\centering
\includegraphics[width=2.97 in]{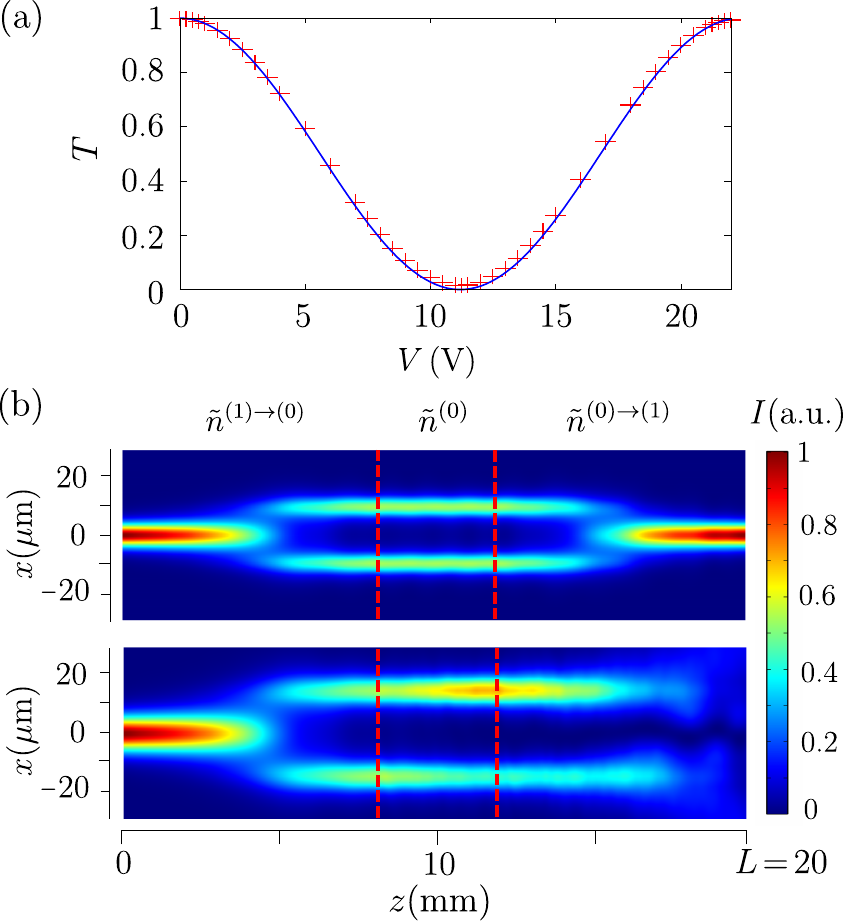}
\caption{(a) Numerically computed transmission (crosses) and theoretical expected curve (solid line) for the MZI as a function of the applied voltage. (b) Numerical simulation of light intensity propagation ($\lambda = 1.55 \, \rm{\mu m}$) along the MZI when the $TE_0^{(1)}$ mode is injected through the input port $\tilde{n}^{(1)}(x)$, and a voltage $V=0$ (upper panel) and $V=11.25 \, \rm{V}$ (lower panel) is applied to the upper arm. The vertical dashed lines delimit the regions of the MZI. Parameter values: $r_{33}\approx 30 \, \rm{pm/V}$, $h=8 \, \mu m$, $l=4 \rm{mm}$ and $\Gamma=0.83$.}
\label{fig:7}
\end{figure}

Finally, we numerically test the performance of the implemented MZI by connecting the $\tilde{n}^{(1)\rightarrow (0)}(x,z)$ and $\tilde{n}^{(0)\rightarrow (1)}(x,z)$ transformations with a central region $\tilde{n}^{(0)}(x)$. When a voltage is applied to one of the arms of the interferometer, a change in the refractive index $\Delta \tilde{n}$ is introduced due to the linear electro-optical effect, introducing a phase difference \cite{bihn2006}:
\begin{equation}
\Delta \phi= k_0 l \Delta \tilde{n}= - k_0 l [\tilde{n}^{(0)}(x)]^3 r_{33} \frac{V}{2h} \Gamma,
\label{eq:phase}
\end{equation}
where $l$ is the electrode length, $\tilde{n}^{(0)}(x)$ is the unperturbed index of refraction, $r_{33}$ is the electro-optical coefficient, $V$ is the applied voltage, $h$ is the distance between the electrodes and $\Gamma$ is the overlap integral between the propagating and the applied electric fields. 

Figure \ref{fig:7}(a) shows the numerically computed transmission $T=I_{out}/I_{in}$ between $-6.7 \, \rm{\mu m} < x < 6.7 \, \rm{\mu m}$ for different voltages (crosses), with the minimum $T_{min}=1.4\%$ at $V=11.25 \, \rm{V}$, which is in good agreement with the expected behavior $T=\cos^2\left(\Delta\phi/2\right)$ (solid line). The MZI has a visibility of $98.6\%$, which could be improved by prolongating the output port $\tilde{n}^{(1)}(x)$ and allowing the radiated mode to propagate away from the waveguide core. We can observe in the upper panel of Fig.~\ref{fig:7}(b) how, if the $TE_0^{(1)}$ mode is injected and no voltage is applied, the beams are recombined at the output port obtaining the $TE_0^{(1)}$ mode with a fidelity $\mathcal{F}_0= 0.998$ for $L=20 \, \rm{mm}$. In the lower panel, we show that if a voltage $V=11.25 \, \rm{V}$ is applied, a phase difference $\Delta \phi = \pi$ is introduced transforming the mode into the $TE_a^{(0)}$ mode, which is radiated during the $\tilde{n}^{(0)\rightarrow (1)}(x,z)$ transformation. 

\section{\label{sec:5}Conclusions}

We have introduced continuous SUSY transformations in systems of optical waveguides, based on the modification of the transverse refractive index profile along the propagation direction. By defining a continuous transformation that connects the superpartner structures, one is able to manipulate the modal content in an adiabatic fashion. 

In particular, we have demonstrated that continuous SUSY transformations offer a systematic way to design efficient and robust (i) tapered waveguides and mode filters by using a single-waveguide structure, and (ii) beam splitters and MZI's by using a two-waveguide structure. Numerically calculated fidelities above 0.999 and above 0.99 have been achieved in a broad region of wavelengths for $L > 1.8 \, \rm{mm}$ tapered waveguides and $L > 6 \, \rm{mm}$ symmetric beam splitters. Moreover, we have also designed a single-waveguide mode filter with fidelities above 0.9 and a MZI with a visibility of $98.6\%$.

As a proof of principle, we have focused on continuous SUSY transformations applied to single- and two-waveguide structures. However, more complex structures could be designed by increasing the number of waveguides, using waveguides with different widths, using optical fibers \cite{Macho2018} or combining continuous with discrete SUSY transformations. Finally, we would like to remark that such transformations are not restricted to optical systems and could be extended to the general formalism of SUSY quantum mechanics and applied to, for instance, trapping potentials modifying its shape in time instead of space \cite{Lahrz2017}.

\section*{Aknowledgments}

The authors gratefully acknowledge financial support through the Spanish Ministry of Economy and Competitiveness (MINECO) (FIS2014-57460-P, FIS2017-86530-P) and the Catalan Government (SGR2017-1646).

\end{document}